\pdfoutput=1

\documentclass[11pt]{article}

\usepackage[]{EACL2023}

\usepackage{times}

\usepackage{times}
\usepackage{latexsym}
\usepackage{amsmath}
\usepackage{amsthm}
\usepackage{booktabs}
\usepackage{multirow}
\usepackage{graphicx}
\usepackage{threeparttable}

\usepackage{enumitem}
\usepackage{float}
\usepackage{subcaption}
\usepackage[T1]{fontenc}
\usepackage{array}

\usepackage[utf8]{inputenc}

\usepackage{microtype}

\title{Extract Class Refactoring Recommendations using Variational Graph Auto-Encoders}

\author{Pritom Saha Akash, Kevin Chen-Chuan Chang \\
    University of Illinois at Urbana-Champaign\\
  \texttt{\{pakash2,kcchang\}@illinois.edu}}

\begin{document}
\maketitle
\begin{abstract}
The code smell is a sign of software design and development flaws reducing the system's reusability and maintainability. Refactoring is done as an ongoing practice to remove the code smell from the program code. Among different code smells, the God class or Blob is one of the most common code smells. A god class contains too many responsibilities, violating object-oriented programming design's low coupling and high cohesiveness principles. This paper proposes an automatic approach to extracting a God class into multiple smaller classes with more specific responsibilities. To do this, we first utilize graph auto-encoder for learning a vector representation for each method (as nodes) in the class after constructing an initial graph. Their structural similarity determines the edge between any two methods. The feature for each method is initialized using different semantic representations. Finally, the learned vectors are used to cluster methods into different groups to be recommended as refactored classes. We assess the proposed framework \footnote{Code and data will be released after review process.} using three different class cohesion metrics on sixteen actual God Classes collected from two well-known open-source systems. A comparative study with a related existing approach shows that the proposed method generates better results for almost all the God Classes. 
\end{abstract}

\section{Introduction}

Code smells are defined as the coding practices that reflect flaws in the design and implementation phases of computer programming. Code smells deteriorate the quality, understandability, and maintainability of a source code in the long run, though they cause no technical inaccuracy. It violates one of the fundamental rules of the Object-Oriented Programming concept (OOP) that a class should focus on a specific responsibility (high cohesion) with limited dependency on other classes (low coupling). However, the system's internal structure undergoes many changes and manipulations during software evolution. For which, the structure of a class may go away from its original design and become complex by incorporating additional responsibilities. This type of code smell is called God class (GC) or Blob \cite{brown1998antipatterns}. As a GC violates the programming principles by increasing coupling and decreasing cohesion within classes, eventually making it difficult for the developers to maintain the system, it is necessary to refactor a GC into smaller, more specific classes. And the refactoring process for this smell is referred to as Extract class refactoring \cite{fowler2018refactoring}.

Several studies have been conducted in the area of extract class refactoring. \cite{bavota2010playing} proposed an approach by following two steps where the method chains in a class are extracted from cohesion within methods. For this, both structural and semantic similarities between methods are considered. Then, using this between-methods similarity matrix, a graph is built, and the edges are cut based on a predefined threshold. A follow-up study was conducted to empirically
evaluate the effectiveness of their tool on real
GCs from existing open-source systems \cite{bavota2014automating}. Another related work based on the MaxFlow-MinCut algorithm was proposed to split a low cohesive class into two classes with higher cohesion. However, this approach can only refactor a GC into two classes. In \cite{bavota2014automating}, an approach for extracting a GC
into multiple classes was proposed by exploiting both structural and semantic similarity between methods. In this work, the semantic similarity between classes was calculated using Latent Semantic Indexing (LSI) \cite{deerwester1990indexing}. However, there is the problem of sparsity in the vector representation of a code snippet (i.e., a method in a class) because of using probabilistic features.

\begin{figure*}[!ht]
\centering
\centerline{\includegraphics[width=\linewidth]{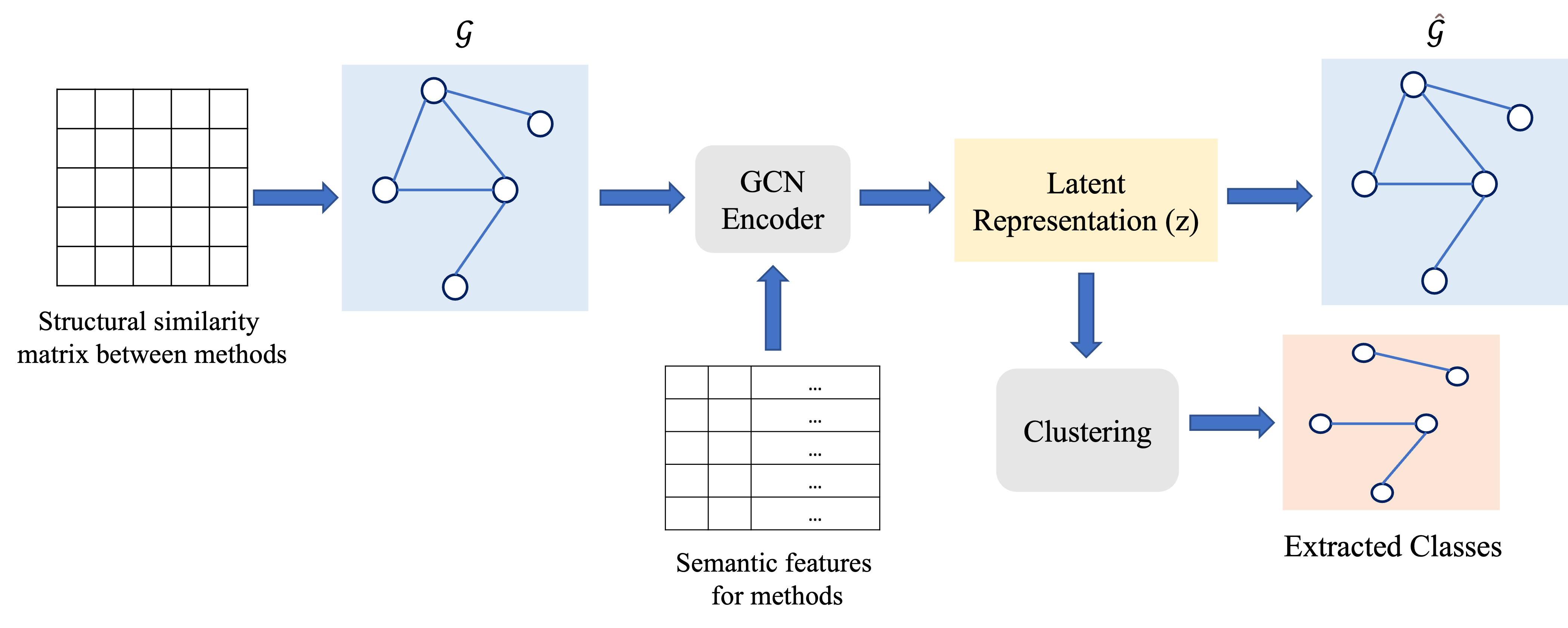}}
\caption{Overview of the proposed architecture}
\label{fig:overview}
\end{figure*}

Different from \cite{bavota2014automating}, recently, \cite{akash2019approach} proposed an approach by using topic modeling where each method in a class is regarded as a document, and a topic distribution is generated using the Latent Dirichlet Allocation (LDA) \cite{blei2003latent} model for each class. Finally, cosine similarity between the two methods' topic distribution is used as their semantic similarity. One major limitation of this method is that the topic model naturally needs a large amount of textual content to train the model. However, a class generally does not have that much textual (both code blocks and in-code documentation) content to train the topic model successfully. Therefore, the learned topic distribution may not be appropriate to represent the semantics of a method in a class (More detailed related work is discussed in Appendix \ref{sec:related_work}). 

On the other hand, one common behavior of the above approaches for refactoring is that they use a simple weighted combination of different similarities between methods. To be more specific, the structural and semantic similarities between the two methods are combined by taking a weighted sum. However, this combination of two different similarities in the same space is not intuitive. There exists a more complex relationship between the structural and semantic similarity between the two methods in the actual scenario. 

Therefore, to solve the problem, we leverage the graph structure of a class consisting of its methods as nodes and edges between them. More specifically, we use the structural similarity within methods to construct the graph topology and explore the different semantic representations of each method to initialize its conceptual attribute. To learn the latent representation of each method in embedding space, we apply a neural graph-based framework called variational graph autoencoder (VGAE) \cite{kipf2016variational}. This model uses a graph-convolutional network (GCN) \cite{kipf2016semi} encoder and a simple inner product decoder to learn the embedding. The GCN can encode both the graph's structure and the nodes' features, thus combining structural and semantic relatedness between methods in a class.

\section{Proposed Framework}
\label{sec:method}
The proposed framework consists of two main steps. The first step is the method-by-method similarity matrix calculation, where each entry of the matrix represents how much proximate two methods are to be included in the same class. The second step is to cluster the methods into different groups based on the calculated similarity matrix in the previous step. The overview of the proposed framework is shown in Figure \ref{fig:overview}.

\subsection{Method-by-Method Similarity Matrix Calculation}
For calculating the method-by-method similarity matrix, we first construct a graph encoding the structural similarity among methods in the concerning class. After that, we initialize a feature representation for each method based on its semantic proprieties. After that, the constructed graph and the feature matrix are used as input to variational graph auto-encoders (VGAE) to learn each method's lower-dimensional latent representation (vector). Finally, the learned vector representation of each method is used to calculate the method-by-method similarity matrix based on cosine similarity. The following sections describe the details of each step.

\begin{figure}
\centering
\centerline{\includegraphics[width=\linewidth]{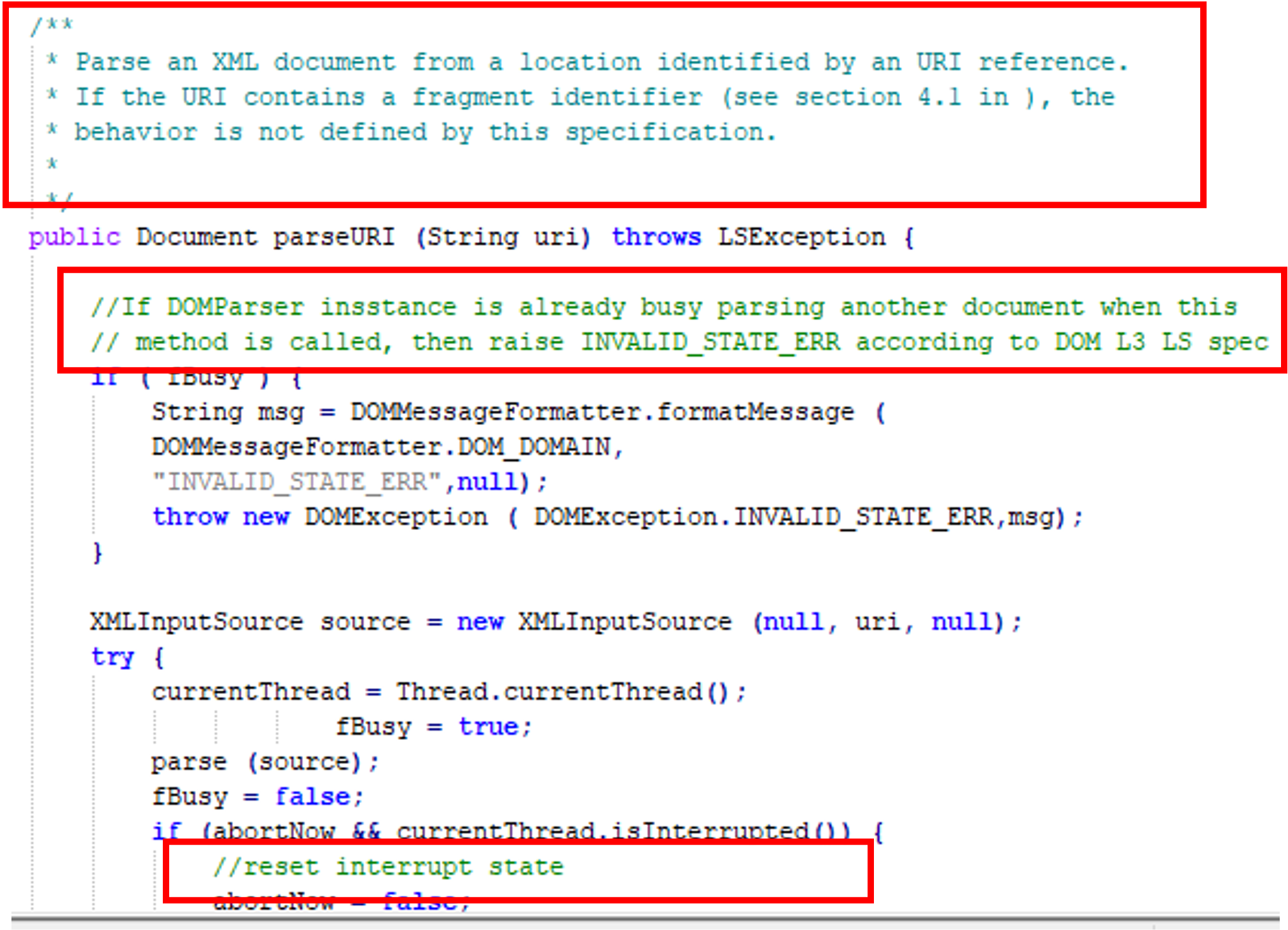}}
\caption{A partial method snippet from DOMParserImpl.java class from an opensource project Xerces}
\label{fig:snippet}
\end{figure}

\subsubsection{Class Graph Construction}
For constructing a graph representing the structure of a class, we need to compute an adjacency matrix. Here, the nodes are methods in the class, and the edge between two nodes represents whether there is a predefined minimum structural similarity between two nodes (i.e., methods). For this, we first need to calculate the structural similarity between the two methods. We use two well-known metrics to capture the structural similarity between methods, and they are Structural Similarity between Methods (SSM) \cite{gui2006coupling}, and Call-based Dependence between Methods (CDM) \cite{bavota2011identifying}.
Let $m_i$ and $m_j$ are two methods, then SSM between these methods is calculated as follows: 
\begin{equation}
    SSM_{i,j} = 
    \begin{cases}
        \frac{\lvert V_i \cap V_j \rvert}{\lvert V_i \cup V_j \rvert } & \text{if } {\lvert V_i \cup V_j \rvert} \neq 0 \\
        0 & \text{otherwise.}
    \end{cases}
\end{equation}
where $V_i$ and $V_j$ refers to the instance variables accessed/used both methods $m_i$ and $m_j$ respectively. The higher value of SSM means two methods are structurally similar to be included in the same class.

On the other hand, the CDM indicates the value of how two methods in a class are related by method calls from each other. Two are structurally proximate if they call each other frequently. The CDM between two methods $m_i$ and $m_j$ is calculated as below:
\begin{equation}
    CDM_{i \rightarrow j} = 
    \begin{cases}
        \frac{calls(m_i, m_j)}{calls_{in}(m_j)} & \text{if } calls_{in}(m_j) \neq 0 \\
        0 & \text{otherwise.}
    \end{cases}
\end{equation}
where $calls(m_i, m_j)$ denotes the number of times method $m_i$ called the method $m_i$ and $calls_{in}(m_j)$ is the total number of incoming calls to method $m_j$. Finally, overall CDM between methods $m_i$ and $m_j$ is calculated as:
    \begin{equation}
        CDM_{i, j} = \text{max}\{ CDM_{i \rightarrow j}, CDM_{j \rightarrow i}\}
    \end{equation}

Now, two structural similarity metrics SSM and CDM are combined with equal weight to determine whether there should be an edge between the two methods. More specifically, if there is a non-zero structural similarity score (combination of SSM and CDM) between the two methods, we draw an edge between these two methods. Therefore, we have a graph constructed on methods as nodes and edges between them by structural similarity. 

\subsubsection{Method Feature Initialization}

Only capturing the structural properties of methods in a class is not enough to thoroughly understand the meaning and characteristics of methods. A method itself contains rich information. For example, as a computer programmer or expert, we can naturally understand what a method is responsible for by looking through its code snippets, including text documentation and comments appearing in it. This textual information is essential to get the semantic meaning of the method and helps understand the relationship between any two methods. An example of a method snippet from an opensource project named \textit{Xerces}\footnote{https://github.com/apache/xerces2-j/tree/trunk/src/org/apache/xerces/parsers} is shown in Figure \ref{fig:snippet}. This figure shows the codes, the text documentation, and comments in a specific method of a class. However, the computer does not understand a text by itself, and for this, we need to represent that in machine-readable vector form. For this purpose, we use this code snippet for each method to turn into a bag of words. Then this bag of words is used to get a vector representation from various embedding methods (described in Section \ref{sec:models}). This vector representation for each method is used as its feature to be utilized in the next step.

\subsubsection{Latent Representation of Methods}
Now, we have a graph of the given GC representing its structural properties with the relation between its methods. We also have a semantic feature for each method calculated using textual information from code snippet of that method. As we said before, we are interested in combining both structural and semantic properties of methods in a class to represent its latent meaning. By doing so, we will be able to calculate the similarity between methods for deciding whether the methods should be extracted to be included in a new refactored class. For this purpose, some previous work \cite{akash2019approach,bavota2014automating,jeba2020god} consider linearly combining structural and semantic similarity using weights. However, this simple linear weighted combination is not appropriate for capturing complex relationships between structural and semantic properties of methods in a method. Rather, considering a class as an attributed network represented by a graph topology constructed by the structural similarity between methods and node attributes calculated considering semantic properties of methods and using this attributed network to learn a latent representation of each method make more sense to encode the complex relationships between two methods in that class. 

For the above purpose, we utilize a recent framework named variational graph autoencoder (VGAE) for unsupervised learning on attributed graph-structured data based on the variational auto-encoder (VAE) \cite{kingma2013auto,rezende2014stochastic}. VGAE learns interpretable latent vector representations of undirected graphs using latent variables. Similar to the original paper, in our study, we use a convolutional graph network (GCN) \cite{kipf2016semi} based encoder to encode the attributed network (graph from structural similarity and semantic feature from previous steps) nodes into a latent lower-dimensional representation and a simple inner product decoder to reconstruct the network from the latent lower-dimensional representation. The overview of the VGAE process is shown in Figure \ref{fig:overview} where we can see the similarity matrix generated adjacency matrix and semantic features of each node are fed into a GCN encoder to get a latent representation matrix $Z$, which is then used to reconstruct the graph based on a Dot Product decoder. Finally, this latent lower-dimensional representation matrix of methods ( $Z$) is used to calculate the method-by-method similarity matrix based on the cosine similarity score.

\subsection{Clustering}

In this step, we use the computed method-by-method similarity matrix to cluster the methods into different groups; each is denoted as a refactored subclass. For this purpose, we use 
OPTICS (Ordering Points To Identify the Clustering Structure) clustering algorithm \cite{ankerst1999optics} to cluster the methods. The first reason for using OPTICS for clustering is that it can find the clusters without pre-specifying the number of clusters beforehand. On the other hand, we can define the minimum number of methods that should be included in a cluster. This requirement is important, while we do not want the extracted class with very few methods.
Finally, the cluster of methods in a class is used as extracted classes as a recommendation to be refactored.

\section{Experimental Setup}
\label{sec:experiment}
\begin{table}[htbp]
\centering
\resizebox{0.9\columnwidth}{!}{
\begin{tabular}{@{}llrr@{}}
\toprule
System                        & God Class                   & LOC                  & \#Methods \\
                              \midrule
\multirow{10}{*}{Xerces}       & AbstractDOMParser           & 2658                & 45                       \\
                              & AbstractSAXParser           & 2418                 & 55                       \\
                              & BaseMarkupSerializer        & 1968                 & 61                       \\
                              & CoreDocumentImpl            & 2871                 & 119                      \\
                              & DeferredDocumentImpl        & 2159                 & 76                       \\
                              & DOMNormalizer               & 2043                 & 31                       \\
                              & DOMParserImpl               & 1393                  & 17                       \\
                              & DurationImpl                & 2021
                               & 45 \\
                              & NonValidatingConfiguration  & 807                  & 18                       \\
                              & XIncludeHandler             & 3082                 & 111                      \\
\midrule
\multirow{7}{*}{GanttProject} & GanttOptions                & 1016                  & 68                       \\
                              & GanttProject                & 1137                 & 90                       \\
                              & GanttGraphicArea            & 366                 & 43                       \\
                              & GanttTaskPropertiesBean     & 707                 & 27                       \\
                              & ResourceLoadGraphicArea     & 1060                 & 29                       \\
                              & TaskImpl                    & 1210                  & 46                       \\
                              \bottomrule
\end{tabular}
}
\caption{God classes used in the experiment}
\label{tab:dataset}
\end{table}
In this section, we empirically evaluate the performance of the proposed framework of extract-class refactoring.
\subsection{Research Questions}
To guide the empirical evaluation of the proposed framework, we seek to answer the following research questions:
\begin{itemize}
    \item (\textbf{RQ1}) Does VGAE improve the results of class refactoring from simple weighted combination-based baselines?

    \item (\textbf{RQ2}) What is the overall best-performing model considering the cohesion and coupling of the newly extracted classes?
    
    \item (\textbf{RQ3}) Do sub-classes after refactoring show better class properties than the original class?
\end{itemize}

\begin{figure*}[!tb]
    \centering
    \subfloat[]{\includegraphics[width=0.25\linewidth]{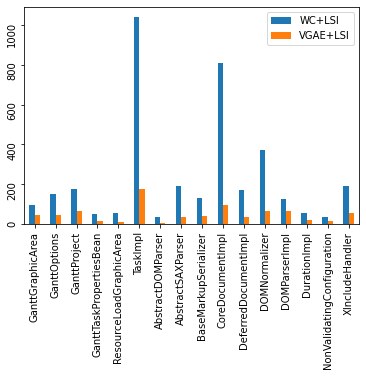}
    \label{fig:lcom_lsi}}
    \subfloat[]{\includegraphics[width=0.25\linewidth]{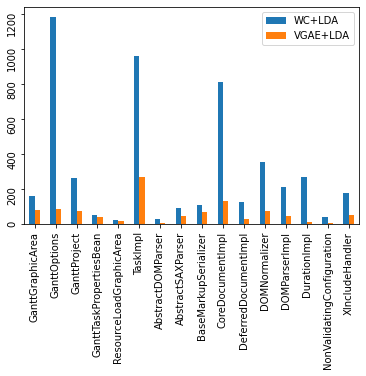}
    \label{fig:lcom_lda}}
    \subfloat[]{\includegraphics[width=0.25\linewidth]{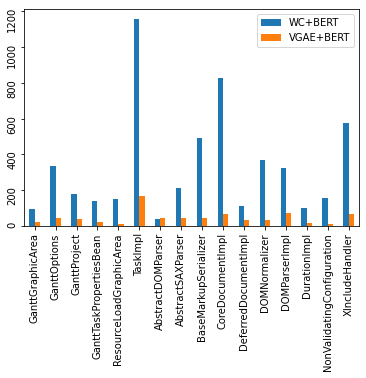}
    \label{fig:lcom_bert}}
    \subfloat[]{\includegraphics[width=0.25\linewidth]{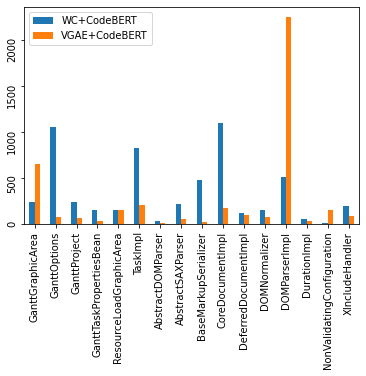}
    \label{fig:lcom_codebert}}
    \caption{Comparative results (average LCOM) between WC and VGAE in 16 classes using four different feature initialization. }
     \label{fig:lcom_result}
\end{figure*}
\begin{figure*}[!tb]
    \centering
    \subfloat[]{\includegraphics[width=0.25\linewidth]{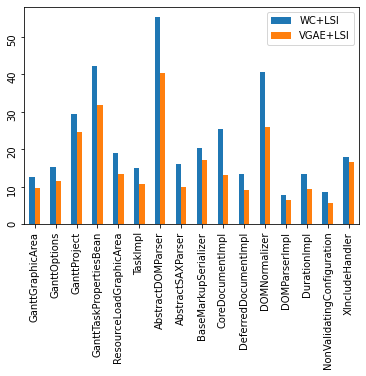}
    \label{fig:mpc_lsi}}
    \subfloat[]{\includegraphics[width=0.25\linewidth]{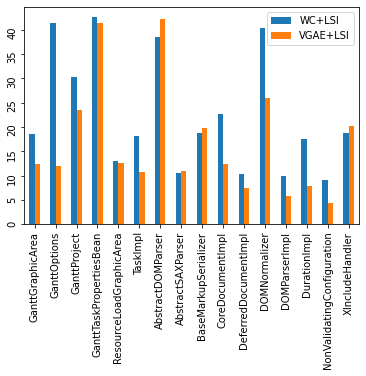}
    \label{fig:mpc_lda}}
    \subfloat[]{\includegraphics[width=0.25\linewidth]{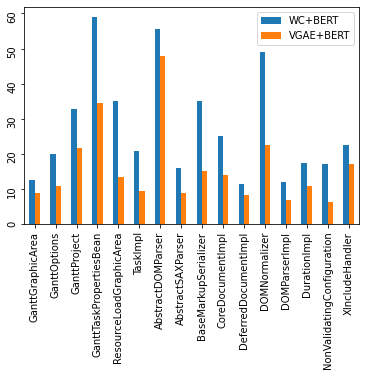}
    \label{fig:mpc_bert}}
    \subfloat[]{\includegraphics[width=0.25\linewidth]{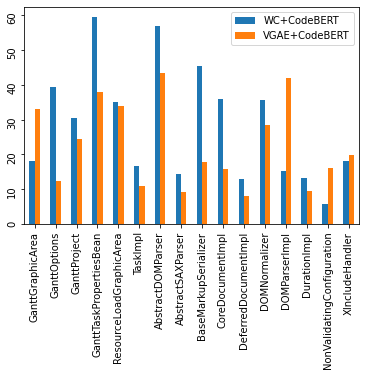}
    \label{fig:mpc_codebert}}
    \caption{Comparative results (average MPC) between WC and VGAE in 16 classes using four different feature initialization.}
     \label{fig:mpc_result}
\end{figure*}
\begin{table*}[htbp]
\centering
\resizebox{0.8\linewidth}{!}{
\begin{tabular}{lrrrrrrrr}
\toprule
Classes                    & \begin{tabular}[c]{@{}r@{}}WC+\\ LSI\end{tabular} & \begin{tabular}[c]{@{}r@{}}WC+\\ LDA\end{tabular} & \begin{tabular}[c]{@{}r@{}}WC+\\ BERT\end{tabular} & \begin{tabular}[c]{@{}r@{}}WC+\\ CodeBERT\end{tabular} & \begin{tabular}[c]{@{}r@{}}VGAE+\\ LSI\end{tabular} & \begin{tabular}[c]{@{}r@{}}VGAE+\\ LDA\end{tabular} & \begin{tabular}[c]{@{}r@{}}VGAE+\\ BERT\end{tabular} & \begin{tabular}[c]{@{}r@{}}VGAE+\\ CodeBERT\end{tabular} \\ \midrule
 
GanttGraphicArea           & 94.33  & 158.5  & 96.67   & 241.0       & 46.5     & 80.33    & \textbf{21.6}      & 656.0         \\
GanttOptions               & 149.67 & 1183.0 & 335.5   & 1059.0      & 43.5     & 85.0     & \textbf{42.62}     & 76.71         \\
 
GanttProject               & 174.86 & 264.17 & 179.67  & 238.33      & 64.33    & 73.44    & \textbf{39.7}      & 70.0          \\
GanttTaskPropertiesBean    & 48.0   & 53.0   & 141.5   & 152.0       & \textbf{17.0}     & 38.67    & 21.0      & 37.0          \\
 
ResourceLoadGraphicArea    & 57.0   & 20.67  & 153.0   & 153.0       & \textbf{11.33}    & 15.33    & 12.0      & 152.0         \\
TaskImpl                   & 1039.6 & 957.2  & 1156.0  & 825.0       & 178.33   & 267.4    & 167.3     & 206.0         \\
 
AbstractDOMParser          & 33.67  & 28.0   & 38.0    & 37.33       & 6.75     & 6.75     & 43.0      & \textbf{6.75}          \\
AbstractSAXParser          & 191.5  & 90.6   & 213.75  & 221.75      & \textbf{33.0 }    & 43.5     & 42.38     & 57.88         \\
 
BaseMarkupSerializer       & 128.8  & 108.0  & 489.5   & 477.5       & 41.17    & 71.0     & 45.0      & \textbf{18.83 }        \\
CoreDocumentImpl           & 808.8  & 810.4  & 826.4   & 1095.67     & 96.92    & 129.9    & \textbf{65.45}     & 175.7         \\
 
DeferredDocumentImpl       & 171.67 & 124.0  & 109.5   & 120.33      & 36.0     & \textbf{26.62}    & 31.5      & 97.33         \\
DOMNormalizer              & 372.0  & 357.0  & 370.0   & 150.0       & 65.0     & 73.83    & \textbf{34.86}     & 71.67         \\
 
DOMParserImpl              & 126.17 & 214.2  & 324.0   & 514.67      & 64.29    & \textbf{46.12}    & 74.86     & 2249.0        \\
DurationImpl               & 53.25  & 270.0  & 99.33   & 58.75       & 18.6     & \textbf{13.6 }    & 15.67     & 37.4          \\
 
NonValidatingConfiguration & 36.5   & 37.5   & 157.0   & 13.0        & 13.0     & \textbf{6.0}      & 13.67     & 156.0         \\
XIncludeHandler            & 188.62 & 177.38 & 574.4   & 201.22      & 55.91    & \textbf{49.1}     & 67.0      & 89.8          \\ \bottomrule
\end{tabular}
}
\caption{Average LCOM results after refactoring (smaller is better)}
\label{tab:result_lcom}
\end{table*}
\begin{table*}[!htb]
\centering
\resizebox{0.8\linewidth}{!}{
\begin{tabular}{lrrrrrrrr}
\toprule
Classes                    & \begin{tabular}[c]{@{}r@{}}WC+\\ LSI\end{tabular} & \begin{tabular}[c]{@{}r@{}}WC+\\ LDA\end{tabular} & \begin{tabular}[c]{@{}r@{}}WC+\\ BERT\end{tabular} & \begin{tabular}[c]{@{}r@{}}WC+\\ CodeBERT\end{tabular} & \begin{tabular}[c]{@{}r@{}}VGAE+\\ LSI\end{tabular} & \begin{tabular}[c]{@{}r@{}}VGAE+\\ LDA\end{tabular} & \begin{tabular}[c]{@{}r@{}}VGAE+\\ BERT\end{tabular} & \begin{tabular}[c]{@{}r@{}}VGAE+\\ CodeBERT\end{tabular} \\ \midrule

GanttGraphicArea           & 12.67  & 18.5   & 12.67   & 18.0        & 9.5      & 12.33    & \textbf{8.8}       & 33.0          \\
GanttOptions               & 15.17  & 41.5   & 20.0    & 39.5        & 11.38    & 11.86    & \textbf{10.75}     & 12.43         \\
 
GanttProject               & 29.29  & 30.33  & 32.67   & 30.5        & 24.67    & 23.56    & \textbf{21.6}      & 24.56         \\
GanttTaskPropertiesBean    & 42.33  & 42.67  & 59.0    & 59.5        & \textbf{31.75}    & 41.33    & 34.5      & 38.0          \\
 
ResourceLoadGraphicArea    & 19.0   & \textbf{13.0}   & 35.0    & 35.0        & 13.33    & 12.67    & 13.33     & 34.0          \\
TaskImpl                   & 15.0   & 18.2   & 20.75   & 16.8        & 10.78    & 10.8     & \textbf{9.3}       & 11.0          \\
 
AbstractDOMParser          & 55.33  & 38.6   & 55.67   & 57.0        & 40.25    & 42.25    & 48.0      & 43.5          \\
AbstractSAXParser          & 16.0   & 10.6   & 16.0    & 14.25       & 9.88     & 11.0     & \textbf{8.75}     & 9.12          \\
 
BaseMarkupSerializer       & 20.4   & 18.8   & 35.0    & 45.5        & 17.0     & 19.8     & \textbf{15.17}     & 17.83         \\
CoreDocumentImpl           & 25.4   & 22.6   & 25.2    & 36.0        & 13.0     & \textbf{12.3}     & 14.09     & 15.7          \\
 
DeferredDocumentImpl       & 13.33  & 10.25  & 11.5    & 13.0        & 9.0      & \textbf{7.38}     & 8.38      & 8.17          \\
DOMNormalizer              & 40.67  & 40.33  & 49.0    & 35.75       & 25.83    & 26.0     & \textbf{22.57}     & 28.33         \\
 
DOMParserImpl              & 7.83   & 10.0   & 12.0    & 15.33       & 6.43     &\textbf{5.75}     & 6.86      & 42.0          \\
DurationImpl               & 13.25  & 17.5   & 17.33   & 13.25       & 9.4      & \textbf{7.8}      & 10.83     & 9.6           \\
 
NonValidatingConfiguration & 8.5    & 9.0    & 17.0    & 5.67        & 5.67     & \textbf{4.25}     & 6.33      & 16.0          \\
XIncludeHandler            & 17.88  & 18.75  & 22.6    & 18.0        & \textbf{16.64}    & 20.3     & 17.1      & 19.7 \\ \bottomrule
\end{tabular}
}
\caption{Average MPC results after refactoring (smaller is better)}
\label{tab:result_mpc}
\end{table*}
\subsection{Dataset}
To evaluate the effectiveness of the proposed framework with existing ones, we applied the framework to refactor the actual sixteen GCs from two well-known open-source systems called Xerces and GanttProject. Table \ref{tab:dataset} summarizes the properties of each class, i.e., the line of codes (LOC) and the number of methods in a class.

\subsection{Implementation Details}
There are some parameters in both baselines and the proposed models we have to set. For example, we use equal weights (1/3) for all three similarities (SSM, CDM, and CSM) to combine two structural and semantic similarities in WC-based baselines. Similarly, in VGAE-based methods, to combine two structural similarities (SSM and CDM) while constructing the class graph, we use equal weights of 0.5. We follow the same pipeline to preprocess text and code in each method for a fair comparison. More specifically, we first tokenize the text in source code based on space, camel case, underscore, special character, and numeric. Then, we remove stop words, including programming language-specific keywords in the source code. After that, we lemmatize each token using Spacy Natural Language Processing \footnote{https://spacy.io/}. Finally, all the tokens are converted to lowercase.

\subsection{Models}
\label{sec:models}
\begin{itemize}
    \item\textbf{WC + LSI.} This baseline is designed following an existing approach of God class refactoring in \cite{bavota2014automating}. This approach combines structural similarity using SSM and CDM with LSI-based semantic similarity with equal weights. For instance, we get three similarity matrices $M_{SSM}$, $M_{CDM}$ and $M_{CSM}$ corresponding SSM, CDM and LSI based CSM. These three matrices are combined with equal weights to get the final method by method similarity matrix $M$. For the calculation of $M_{CSM}$, each method in the class is represented by an LSI-generated vector using the code snippet of that method alongside comments and code documentation. Finally, the similarity $M$ is used to cluster the methods into multiple classes.
    \item \textbf{WC + LDA.} This baseline is also an existing approach \cite{akash2019approach}. It has a similar overall framework to the previous baseline. Rather than using LSI, this baseline uses LDA for semantically representing each method in a class. More specifically, this method uses textual information from the methods to find a set of topics and their distribution using the LDA topic modeling algorithm \cite{blei2003latent} by considering each method as a document. Then, the topic distribution for each method is used as the semantic representation of that method and used to calculate the cosine similarity between any two methods.
    
    \item \textbf{WC + BERT.} This baseline also uses the weighted combination of structural and semantic similarity. For calculating semantic similarity, each method is represented in a vector space using sentence BERT \cite{reimers2019sentence}. Then cosine similarity between two corresponding vectors is calculated.
    \item \textbf{WC + CodeBERT.} It is precisely similar to the WC+BERT model, except instead of using BERT for method representation, CodeBERT \cite{feng2020codebert} is used to get vector representation of each method in the class. CodeBert is a pre-trained language model specified for programming codes and is used to encode a code snippet (i.e., a method in a class) into a vector space. 
    \item \textbf{VGAE + LSI.} This is the proposed framework where the structural similarity combining SSM and CDM is used to construct the class graph. For the initial node feature, each method's LSI-based vector representation is used. Then, VGAE \cite{kipf2016variational} is used to learn a latent semantic representation of each method. Finally, the learned vectors of methods are used to cluster them into groups that are recommended as refactored classes. 
    \item \textbf{VGAE + LDA.} This proposed framework is similar to the previous one. However, unlike the previous one, in this approach, instead of using LSI-based vector representation for the initial node feature, we used LDA-based vector representation (the same representation used in WC+LDA).
    \item \textbf{VGAE + BERT.} In this VGAE-based framework, we use vector representation of each method by BERT pre-trained language model as initial node feature representation (same representation used in WC+BERT). The rest of the framework is similar to the previous one.
    \item \textbf{VGAE + CodeBERT.} In this VGAE-based framework, we use vector representation of each method by CodeBERT pre-trained language model for programming language code as initial node feature representation (same representation used in WC+CodeBERT). The rest of the framework is similar to the previous one.
\end{itemize}

\subsection{Evaluation Metrics}
To assess the performance of proposed models, we use two evaluation metrics named Lack of Cohesion of Methods (LCOM) \cite{li1993maintenance} and Message Passing Coupling (MPC) \cite{li1993maintenance}. LCOM counts the difference between the number of pairs of methods that do not share any instance variables and the number of pairs where two methods share at least one instance variable. Hence, the lower value of LCOM is better and means there is high cohesion in the class. On the other hand, MPC measures the number of method calls defined in methods of a class to methods in other classes. Again, a lower MPC is better because the class should have lower coupling with other classes according to the object-oriented design principle.

\section{Results and Discussion}
\subsection{Performance of VGAE over Weighted Combining (WC)}
\textbf{RQ1:} To answer the first research question, we compare the average LCOM and average MPC scores of extracted classes over 16 GCs between WC and VGAE-based methods. 

Figures \ref{fig:lcom_result} and \ref{fig:mpc_result} show the result of LCOM and MPC respectively. From figure \ref{fig:lcom_result}, for the cases of almost all of the 16 GCs and four different feature initialization, we can see the average LCOM scores over the extracted classes by VGAE-based methods are far lower than that of WC-based methods. This result means that VGAE-based methods extract GCs to sub-classes with more cohesiveness among the methods in the sub-classes compared to the simple WC-based methods. 

Similarly, from figure \ref{fig:mpc_result}, we can observe that the average MPC score over the extracted classes by VGAE-based methods is considerably lower than that of WC-based methods in almost all of the cases. It means that the coupling among the extracted classes by VGAE-based methods is lower than that of sim WC-based methods.

As per the object-oriented design principles, we know that it is desirable to have higher cohesion among methods in a class and lower coupling among classes. From the above result, we can see that the VGAE-based methods can achieve higher cohesion and lower coupling among the extracted classes than the simple WC-based methods. Therefore, we can easily say that the use of VGAE to combine structural and semantic properties of methods in a class has considerable improvement over the simple weighted combination in terms of performance.
\begin{table*}[!htb]
\centering
\resizebox{0.9\linewidth}{!}{%
\begin{tabular}{llrrcrr}
\toprule

\multirow{2}{*}{System}                          & \multirow{2}{*}{Class Names} & \multicolumn{2}{l}{\begin{tabular}[c]{@{}l@{}}Before \\ Refactoring\end{tabular}} & \multicolumn{3}{c}{After Refactoring}                                                                                                                                                                                               \\ \cmidrule(l){3-4} \cmidrule(l){5-7} 
                                                  &                                                     & LCOM                        & MPC                      & \begin{tabular}[c]{@{}l@{}}\#splits\end{tabular}                                                                & LCOM                                                       & MPC                                 \\ \midrule
\multicolumn{1}{c}{\multirow{6}{*}{GanttProject}} & GanttGraphicArea                                 & 657.0                       & 34                       & 5                                                                                                   & 64.0, 10.0, 0, 13.0, 21.0                                  & 12, 7, 11, 2, 12                    \\
\multicolumn{1}{c}{}                              & GanttOptions                                    & 2626.0                      & 62                       & 8                                                                                  & 218.0, 26.0, 8.0, 20.0, 9.0, 9.0, 12.0, 39.0               & 64, 0, 17, 0, 4, 0, 0, 1            \\
\multicolumn{1}{c}{}                              & GanttProject                                      & 3632.0                      & 147                      & 10                                                                   & 323.0, 16.0, 0, 21.0, 0, 9.0, 15.0, 3.0, 10.0, 0           & 45, 32, 6, 8, 26, 14, 16, 47, 4, 18 \\
\multicolumn{1}{c}{}                              & GanttTaskPropertiesBean                         & 419.0                       & 97                       & 4                                                                                                        & 13.0, 0, 0, 71.0                                           & 68, 23, 2, 45                       \\
\multicolumn{1}{c}{}                              & ResourceLoadGraphicArea                           & 153.0                       & 35                       & 3                                                                                                            & 8.0, 28.0, 0                                               & 7, 13, 20                           \\
\multicolumn{1}{c}{}                              & TaskImpl                                        & 7491.0                      & 48                       & 10                                                                       & 1568.0, 6.0, 10.0, 15.0, 36.0, 0, 30.0, 0, 0, 8.0          & 38, 5, 0, 0, 9, 3, 11, 1, 26, 0     \\ \midrule
Average                                           &                                                 & 2496.33                     & 70.50                    &                                                                                                                            & 65.98                                                      & 15.43                               \\ \midrule
\multirow{10}{*}{Xerces}                          & AbstractDOMParser                                & 4.0                         & 130                      & 4                                                                                                        & 172.0, 0, 0, 0                                             & 27, 47, 48, 70                      \\
                                                  & AbstractSAXParser                                & 2308.0                      & 38                       & 8                                                                               & 339.0, 0, 0, 0, 0, 0, 0, 0                                 & 28, 2, 18, 1, 7, 7, 3, 4            \\
                                                  & BaseMarkupSerializer                              & 1004.0                      & 64                       & 6                                                                                         & 250.0, 10.0, 10.0, 0, 0, 0                                 & 64, 0, 1, 3, 15, 8                  \\
                                                  & CoreDocumentImpl                                 & 6589.0                      & 76                       & 11                                                             & 333.0, 0, 0, 20.0, 0, 66.0, 151.0, 10.0, 10.0, 120.0, 10.0 & 48, 7, 4, 24, 4, 44, 19, 4, 0, 1, 0 \\
                                                  & DeferredDocumentImpl                             & 987.0                       & 20                       & 8                                                                                & 235.0, 7.0, 10.0, 0, 0, 0, 0, 0                            & 19, 2, 12, 5, 9, 17, 2, 1           \\
                                                  & DOMNormalizer                                  & 1729.0                      & 102                      & 7                                                                                      & 174.0, 7.0, 10.0, 4.0, 0, 28.0, 21.0                       & 39, 82, 3, 23, 11, 0, 0             \\
                                                  & DOMParserImpl                                    & 2250.0                      & 43                       & 7                                                                                      & 433.0, 4.0, 10.0, 10.0, 55.0, 9.0, 3.0                     & 8, 16, 0, 0, 0, 0, 24               \\
                                                  & DurationImpl                                    & 716.0                       & 23                       & 6                                                                                           & 27.0, 0, 0, 21.0, 36.0, 10.0                               & 30, 9, 11, 2, 9, 4                  \\
                                                  & NonValidatingConfiguration                       & 157.0                       & 17                       & 3                                                                                                             & 28.0, 2.0, 11.0                                            & 6, 3, 10                            \\
                                                  & XIncludeHandler                                  & 4790.0                      & 83                       & 10                                                                    & 639.0, 0, 0, 11.0, 1.0, 0, 0, 19.0, 0, 0                   & 81, 3, 5, 5, 33, 8, 18, 8, 10, 0    \\ \midrule
Average                                           &                                                 & 2053.4                      & 59.6                     &                                                                                                                          & 47.51                                                      & 14.8                                \\ \bottomrule
\end{tabular}}

\caption{Cohesion (LCOM) and Coupling (MPC) results obtained before and after refactoring the 16 God Classes.}
\label{tab:result_rq3}
\end{table*}
\subsection{Overall Best Performing Model}
\textbf{RQ2:} 
To answer the second research question, we compare the results of six models over 16 GCs using the two metrics, LCOM and MPC. More specifically, we calculate the average LCOM and MPC over the extracted classes for each of the sixteen GCs by six compared models.

Table \ref{tab:result_lcom} shows the comparative results using average LCOM score for six models. From the Table, we can see that almost all the best-performing models in average LCOM are from the VGAE-based models. Among VGAE-based models, no single method marginally outperforms others in most of the sixteen GCs. However, considering overall LCOM scores, we can see the VGAE+BERT and  VGAE+LDA models attain the lowest average LCOM for 5 out of 16 GCs by each.
On the other hand, Table shows the comparative results using the average MPC score for six models. In this Table, we again see the best-performing models are the VGAE-based models. Among the VGAE-based models, we observe that the VGAE+BERT obtains lower average MPC scores in 7 out of 16 GCs, where the VGAE+LDA is the second-best performing model in terms of average MPC score.  

Therefore, considering both LCOM and MPC, we can say that overall, the VGAE+BERT model is the best-performing model to extract GCs into sub-classes with higher cohesion and lower coupling. Interestingly, the VGAE+CodeBERT model does not achieve as good a performance in extracting GCs as the VGAE+BERT model. The initial motivation for using CodeBERT was that as CodeBERT is a pre-trained LM dedicated to programming source code, it will be better to encode the semantic meaning of source code alongside syntactic information than BERT. However, this intuition may not be correct as we can not see the reflection of that intuition in the result. The reason is that while representing a method semantically, the only important thing that matters is the text in the code rather than its syntax. Therefore, BERT is better at capturing the semantic meaning of a method than CodeBERT.

\subsection{Improvement over Original Class}
\textbf{RQ3:} To answer the third research question, we compare the result of LCOM and MPC on each of the original 16 GCs before and after refactoring by the best-performing model VGAE+BERT. The result is shown in Table \ref{tab:result_rq3}. From the Table, it is clearly observable that for all the 16 GCs, the cohesion is increased among extracted classes after refactoring(by lower LCOM), and the coupling (by lower MPC) decreases among the extracted classes after refactoring. It indicates that the proposed refactoring model for extracting GCs is effective in finding sub-classes with specific responsibilities. For instance, the LCOM and MPC values of \textit{GanttGraphicArea} were 657 and 34, respectively. The proposed approach refactors the class into five sub-classes. Each of the LCOM scores (64, 10, 0, 13, and 21) of all newly generated classes is substantially lower than the original class. Similarly, each of the MPC scores of newly generated classes is also lower than the original one (12,7,11,2 and 12). Therefore, this result indicates that the sub-classes after refactoring using the proposed method show better class properties than the original one. 

\section{Conclusion}
This paper proposes a new framework for refactoring a God class into sub-classes. After refactoring using the proposed framework, the extracted sub-classes obtain higher cohesion and lower coupling. 
The proposed approach combines the structural and semantic properties of methods in a God class based on the variational graph auto-encoder (VGAE) model. We conducted an empirical experiment over 16 actual GCs from two open-source systems to evaluate the performance of the proposed framework. The experimental result shows that the proposed framework outperforms the existing and well-designed baseline with better cohesion and coupling score.

\section*{Limitations}
One limitation of the proposed study is that it can only give suggestions of extracting a predefined GC into sub-classes. It can’t detect whether a class is a God class or not. As a future research direction, we plan to propose an automatic approach to detecting and extracting GCs.


 
\bibliography{anthology}
\bibliographystyle{acl_natbib}

\appendix

\section{Related Work}
\label{sec:related_work}
Several works are conducted in the extract-class refactoring task. The existing literature on this area reflects some patterns. For instance, most previous studies in extract-class refactoring focus on addressing the problem of low cohesion and high coupling in god class by splitting the class into smaller classes. More specifically, these studies refactor the god class into sub-classes such that those classes are loosely coupled and highly cohesive. Therefore, these methods are mainly metric-based refactoring approaches. 

For example, a series of extract-class refactoring methods was proposed \cite{bavota2010playing,bavota2011identifying,bavota2014automating} where the main focus was on identifying method chains for recommending new classes possessing higher cohesion than the initial class. These methods calculate cohesion between two methods in the class, combining both structural and semantic similarity measures. After that, the authors constructed a weighted graph of methods where edges were represented by the calculated cohesion score between the two methods. These approaches used different graph-cutting mechanisms in different studies for clustering methods into multiple groups. For instance, in \cite{bavota2011identifying},  the authors used the max-flow min-cut algorithm to divide the graph into two subgraphs, each representing a new subclass. However, the main limitation of this approach is that it can refactor a god class into only two classes which may not be practical for many actual god classes. 

To support multi-class refactoring, the authors extended their work by using a two-step mechanism to cut the graph into multiple subgraphs based on a predefined cohesion threshold (edge weight) \cite{bavota2010playing,bavota2014automating}. As said before, these methods used both structural and semantic similarities between methods. For structural similarity, they used two well-known metrics named Structural Similarity be-
tween Methods (SSM) \cite{gui2006coupling}, and
Call-based Dependence between Methods (CDM) \cite{bavota2011identifying}. For semantic similarity, they first used textual information in source code to get a vector representation of each method using Latent Semantic Indexing (LSI) \cite{deerwester1990indexing}. Then, these vectors are used to calculate the cosine similarities between methods. Recently, a similar approach \cite{akash2019approach} follows a similar framework of combining both structural and semantic similarities between methods. However, instead of using LSI to represent a method semantically, this approach applies a topic modeling algorithm called Latent Dirichlet Allocation (LDA) \cite{blei2003latent} on source code to represent a method by a distribution of topics.

\end{document}